\documentclass[draft]{agujournal2019}
\usepackage{url} 
\usepackage{lineno}
\usepackage[finalnew]{trackchanges}

\usepackage{soul}
\usepackage[version=3]{mhchem}
\draftfalse

\journalname{JGR: Planets}

\usepackage{pdfpages}

\begin{document}


\title{Three-Dimensional Venus Cloud Structure Simulated by a General Circulation Model}

%
%

\authors{Wencheng D. Shao\affil{1}, Jo\~ao M. Mendon\c ca\affil{1}, Longkang Dai\affil{2}}

\affiliation{1}{National Space Institute, Technical University of Denmark, Lyngby, Denmark}
\affiliation{2}{College of Meteorology and Oceanography, National University of Defense Technology, Changsha, China}

\correspondingauthor{Wencheng D. Shao}{wenshao@dtu.dk}


\begin{keypoints}
\item We construct a Venus climate model with cloud physics, and the cloud vertical structure agrees with observations.
\item \ce{H2SO4} and \ce{H2O} vapors in the middle cloud basically follow their SVMRs and show higher concentrations at low latitudes.
\item The semidiurnal thermal tide affects \ce{H2SO4} and \ce{H2O} vapors, cloud mass loading and acidity at different altitudes.
\end{keypoints}

%
%

%
%


\begin{abstract}
The clouds have a great impact on Venus's energy budget and climate evolution, but its three-dimensional structure is still not well understood. Here we incorporate a simple Venus cloud physics scheme into a flexible GCM to investigate the three-dimensional cloud spatial variability. Our simulations show good agreement with observations in terms of the vertical profiles of clouds and \ce{H2SO4} vapor. \ce{H2O} vapor is overestimated above the clouds due to efficient transport in the cloud region. The cloud top decreases as latitude increases, qualitatively consistent with Venus Express observations. The underlying mechanism is the combination of \mbox{\ce{H2SO4}} chemical production and meridional circulation. The mixing ratios of \ce{H2SO4} at 50-60 km and \ce{H2O} vapors in the main cloud deck basically exhibit maxima around the equator, due to the effect of temperature's control on the saturation vapor mixing ratios of the two species. The cloud mass distribution is subject to both \ce{H2SO4} chemical production and dynamical transport and shows a pattern that peaks around the equator in the upper cloud while peaks at mid-high latitudes in the middle cloud. At low latitudes, \ce{H2SO4} and \ce{H2O} vapors, cloud mass loading and acidity show semidiurnal variations at different altitude ranges, which can be validated against future missions. Our model emphasizes the complexity of the Venus climate system and the great need for more observations and simulations to unravel its spatial variability and underlying atmospheric and/or geological processes.
\end{abstract}

\section*{Plain Language Summary}
On Venus, highly reflective clouds cover the surface entirely. This means that the clouds greatly impact Venus's current and, very likely, past energy budget. Therefore, understanding the Venus clouds is essential to constructing a full paradigm of the Venus climate and evolution. However, due to the lack of both three-dimensional, long-term observations and comprehensive climate models, the cloud spatial structure and its impact on atmospheric processes remain elusive. Here, we construct a three-dimensional climate model that includes cloud physics and simple chemistry as the first step toward fully understanding the Venus clouds. Our simulated vertical profiles of the Venus clouds agree well with observations. We find that the condensable gases, sulfuric acid and water vapors in the cloud region become more abundant in the lower latitudes due to the temperature difference over different latitudes. The cloud top becomes lower as it approaches the polar region, and the underpinning processes are related to sulfuric acid chemical production and meridional circulation. The equatorial cloud structure shows semidiurnal features, which are related to the excited thermal tides in the Venus atmosphere. Our study is preparing for future Venus missions like EnVision, to maximize their science returns.

\section{Motivation}
Venus may have formed with similar compositions to Earth but has since evolved in a distinct path that has resulted in its current inhospitable climate. To fully understand its climate system, it is crucial to investigate the chemistry-cloud-dynamics coupling on this planet.  Clouds on Venus significantly influence its climate evolution. For instance, clouds could have formed a cold trap for water and reduce the loss of hydrogen in the Venus history \cite{bullock2001recent} ; asymmetry clouds may have formed on the day and night sides of Venus at its early age and exert a warming effect on this planet \cite{Turbet2021} . Clouds are also an crucial component of the current Venus climate system. Clouds are closely connected to volatile cycles (e.g., sulfur cycle) on this planet \cite<e.g.>{bierson2020chemical, krasnopolsky2015vertical, dai_simple_2022}  and thus can indicate surface-atmosphere volatile exchanges (e.g., volcanism). Clouds also act as the boundary between the lower and upper atmosphere, and thus is important for understanding mass, momentum, energy exchange (e.g. \mbox{\ce{SO2}} and \mbox{\ce{H2O}} depletion in the cloud region) between the two parts.   In the upper cloud, sulfuric acid clouds reflect about three quarters of the incident solar radiation, and the unkown UV absorber in the clouds contributes to almost half of the solar energy deposited on this planet \cite{crisp1986radiative} . Thus, cloud spatial heteorogeneity is essential for studying energy budget and atmospheric circulations on Venus. To fully understand the Venus climate, a well-established framework of the three-dimensional cloud structure is inevitable. This requires an amount of effort from both observational, experimental and theoretical scientists. This is especially urgent at this time since at least three Venus missions (EnVision, DAVINCI+ and VERITAS) are going to visit  this planet within a decade.

Remote sensing of Venus has been challenging due to its thick atmosphere and cloud deck ($\sim47-70\,$km). As a result, the upper part ($>60\,$km) of the Venus atmosphere is currently the most observed and studied region \cite<e.g.,>{vandaele2017sulfur, vandaele2017sulfurii, marcq2020climatology}. The mesosphere ($\sim60-100\,$km) is characterized by a complicated photochemical network \cite{yung1982photochemistry, mills1998observations, zhang2012sulfur, krasnopolsky2012photochemical, krasnopolsky2018disulfur, bierson2020chemical, rimmer2021hydroxide}. The sulfur photochemistry in the mesosphere is critical to the upper cloud ($\sim58-70\,$km) formation, while the lower and middle clouds ($\sim47-58\,$km) rely more on convective activity \cite{titov2018clouds, dai_simple_2022}. Venus Express (VEx) and Akatsuki both provide abundant cloud observations that show cloud morphological features \cite<e.g.,>{titov2012morphology, peralta2019new, limaye2018venus}. Generally, the cloud top decreases from low and middle latitudes to high latitudes \cite{2014Icar..232..232H}. In the equatorial region, clouds are more patchy and spatially variable, possibly related to convective activities and waves \cite{2008Natur.456..620T, narita2022correlation}. At middle and high latitudes, cloud spatial patterns are smoother and characterized by streaks and bands, suggesting convectively stable layers there \cite{2008Natur.456..620T}. In the polar region, planetary vortices are observed at infrared wavelengths \cite{2008Natur.456..620T, garate2013chaotic}. However, the three-dimensional (3D) cloud structure and its temporal variability are still not well understood.

To fully understand the Venus clouds, a 3D GCM that fully couples dynamics, clouds, photochemistry and radiative transfer is needed. This is because the clouds on Venus are closely related to photochemistry in the mesosphere, and cloud radiative feedback greatly impacts the Venus climate system. There has been some modelling work on understanding the Venus cloud features. 1D microphysical models \cite{imamura2001microphysics, mcgouldrick2007investigation, gao2014bimodal, parkinson2015distribution, karyu2024one} have been used to study the vertical profiles of the Venus clouds, including cloud mass, particle size distribution, and even the temporal and spatial variabilities of the Venus upper haze \cite{parkinson2015distribution}. On the other side, \citeA{krasnopolsky2015vertical} and \citeA{dai_simple_2022} both investigated the vertical profiles of the Venus clouds by applying a simple cloud physics model, without detailed microphysical processes. For two dimensions, \citeA{imamura1998venus} used a 2D cloud model to investigate how meridional circulation affects the Venus cloud distributions. They found that meridional circulation can produce large cloud mass at low and high latitudes, consistent with observations.

The three-dimensional studies have all highlighted the importance of meridional circulation and other atmospheric dynamical processes for the Venus clouds. \citeA{lee2010bulk} included a simple bulk cloud parameterization scheme in a Venus GCM and showed that the ``Y'' shaped cloud structure can be produced by dynamics alone. \citeA{ando2020venusian} incorporated a simple cloud physics scheme into a Venus GCM and indicated that atmospheric waves, disturbances, and mean meridional circulation can have strong effects on the cloud structure. \citeA{stolzenbach2023three} presented the Venus GCM that first includes both photochemistry and clouds, and investigated the importance of the Hadley-type circulation to long-lived species. \citeA{karyu2023vertical} incorporated a simple cloud parameterization into a Venus GCM and studied the effects of gravity and Kelvin waves on cloud opacity variations. Due to the computational cost of accurately simulating cloud physics in 3D, all the above 3D Venus cloud models use a rather simple cloud representation, which is also the case in this work.

In this study, we investigate the three-dimensional cloud structure on Venus by implementing the cloud physics scheme from \citeA{dai_simple_2022} that resolves cloud acidity self-consistently into a state-of-the-art Venus GCM, OASIS, by \citeA{mendoncca2020modelling}. The cloud scheme and the GCM have demonstrated their capability of simulating the Venus clouds and dynamics, respectively. Section 2 will describe the methodology we use, and Section 3 will present the simulated 3D cloud structure. We discuss the 3D structure in vertical (Section 3.1), zonal (Section 3.2) and local-time (Section 3.3) dimensions. In Section 4, we conclude our work and discuss the potential future improvements of our model.

\section{Model Description}
\label{sec2}
We construct a new cloudy GCM for Venus. The GCM framework is a state-of-the-art Venus GCM, OASIS, developed by \citeA{mendoncca2020modelling}. OASIS is a flexible planetary laboratory developed upon a dynamical core, THOR \cite{mendoncca2016thor,deitrick2020thor}, and has been used to study the Venus atmosphere as well as exoplanetary atmospheres \cite<e.g.,>{mendoncca2018chemistry, mendoncca2018revisiting,Deitrick2022}. The model \change{solves}{can solve} the three-dimensional non-hydrostatic Euler equations, \add{but here for simplicity we assume hydrostatic equilibrium for the Venus atmosphere. The model} uses an icosahedral grid and bases its computations mainly on Graphic Processing Unit (GPU; currently only applicable on Nvidia GPUs). The model's grid allows for accurate simulations of the polar-region dynamics with high computational efficiency since it has cells with similar shapes and areas everywhere and thus does not have heavy distortions or singularities in the polar region that many other grids have. OASIS has modules for dynamics, radiation, soil, turbulence, chemistry and clouds \cite{mendoncca2020modelling}. In this study, we use the radiation scheme from \citeA{mendoncca2015new} and the cloud physics from \citeA{dai_simple_2022}. The soil and turbulence schemes are the same as those in \citeA{mendoncca2020modelling}; we assume a flat basalt surface for Venus and use a soil formulation scheme similar to that of LMD Mars GCM \cite{1993JAtS...50.3625H} and of Oxford Venus GCM \cite{mendoncca2016exploring}; the turbulence scheme here is a fourth-order hyperdiffusion operator coupled with a 3D divergence damping, to avoid the accumulation of high-frequency waves at the smallest scale resolved by the numerical model.

The cloud scheme includes four species, the gas and liquid phases of \ce{H2SO4} and \ce{H2O}. It resolves cloud condensation/evaporation and cloud sedimentation. The advantage of using this scheme is that it resolves the dependence of \ce{H2SO4} and \ce{H2O} saturation vapor pressures on cloud acidity (defined as \ce{H2SO4} weight percent here) and requires less computational resource than detailed-microphysics models. The condensation rates of \ce{H2SO4} and \ce{H2O} are expressed as (Formula 9 and 10 in \citeA{dai_simple_2022}):
\begin{equation}
\begin{aligned}
S_{cond}=\frac{2\pi n_{p}D_{1}M_{1}f_{m_{1}}n_{atm}D_{p}}{M_{1}+M_{2}m} \left(q_{1}^{g}-q_{1}^{svp}\right),
\label{eq1}
\end{aligned}
\end{equation}
and
\begin{equation}
\begin{aligned}
S_{cond}=\frac{2\pi n_{p}D_{2}M_{2}f_{m_{2}}n_{atm}D_{p}}{\frac{M_{1}}{m}+M_{2}} \left(q_{2}^{g}-q_{2}^{svp}\right),
\label{eq2}
\end{aligned}
\end{equation}
where $n_p$ is the CCN number density, $n_{atm}$ the total atmospheric number density, $D_p$ the particle diameter and $m$ the \ce{H2O}/\ce{H2SO4} molecular ratio in the cloud droplet. $D_i$ is the molecular diffusion coefficient, $M_i$ the relative molecular weight, and $f_{m_i}$ the flux-matching factor, with i=1 for \ce{H2SO4} and i=2 for \ce{H2O}. $q_{i}^{g}$ is the volumn mixing ratio for vapor i, and $q_{i}^{svp}$ is the saturation vapor mixing ratio (SVMR) for vapor i. SVMR is converted from the saturation vapor pressure (SVP). SVP is calculated using formula 11 in \mbox{\citeA{dai_simple_2022}}, which has considered SVP above pure condensed species \cite{kulmala1990binary, tabazadeh1997new} , chemical potential change effect \mbox{\cite{zeleznik1991thermodynamic}} and Kelvin effect \mbox{\cite{seinfeld2016atmospheric}}. Cloud acidity affects condensation rate through $m$ and $q_{i}^{svp}$ in formula (1-2). \add{Note that the SVP formula for \mbox{\ce{H2SO4}} from} \citeA{kulmala1990binary} \add{has been investigated over the temperature range of about 153-363 K, which is relevant for the Venus cloud region. The SVP formula for \mbox{\ce{H2O}} from} \citeA{tabazadeh1997new} \add{is derived in the temperature range of 185-260 K, but we have compared its extrapolation over 150-400 K with other works} \cite{murphy2005review, nachbar2019vapor} \add{that apply to different temperature ranges (not shown here). We find this \mbox{\ce{H2O}} SVP formula is also applicable to 150-400 K.}

The sedimentation rate is expressed as (Formula 19 in \citeA{dai_simple_2022}):
\begin{equation}
\begin{aligned}
S_{sed, i}=-\frac{\partial}{\partial z} \left(v n_{atm} q_{i}^{l}\right),
\label{eq3}
\end{aligned}
\end{equation}
where $v$ is the Stokes velocity, and $q_i^l$ is the mixing ratio for liquid species i (i=1 for \ce{H2SO4}, and i=2 for \ce{H2O}). The parameters used to calculate these condensation and sedimentation rates are adopted as the same as those in \citeA{dai_simple_2022}.

\begin{figure}
\noindent\includegraphics[width=1.0\columnwidth]{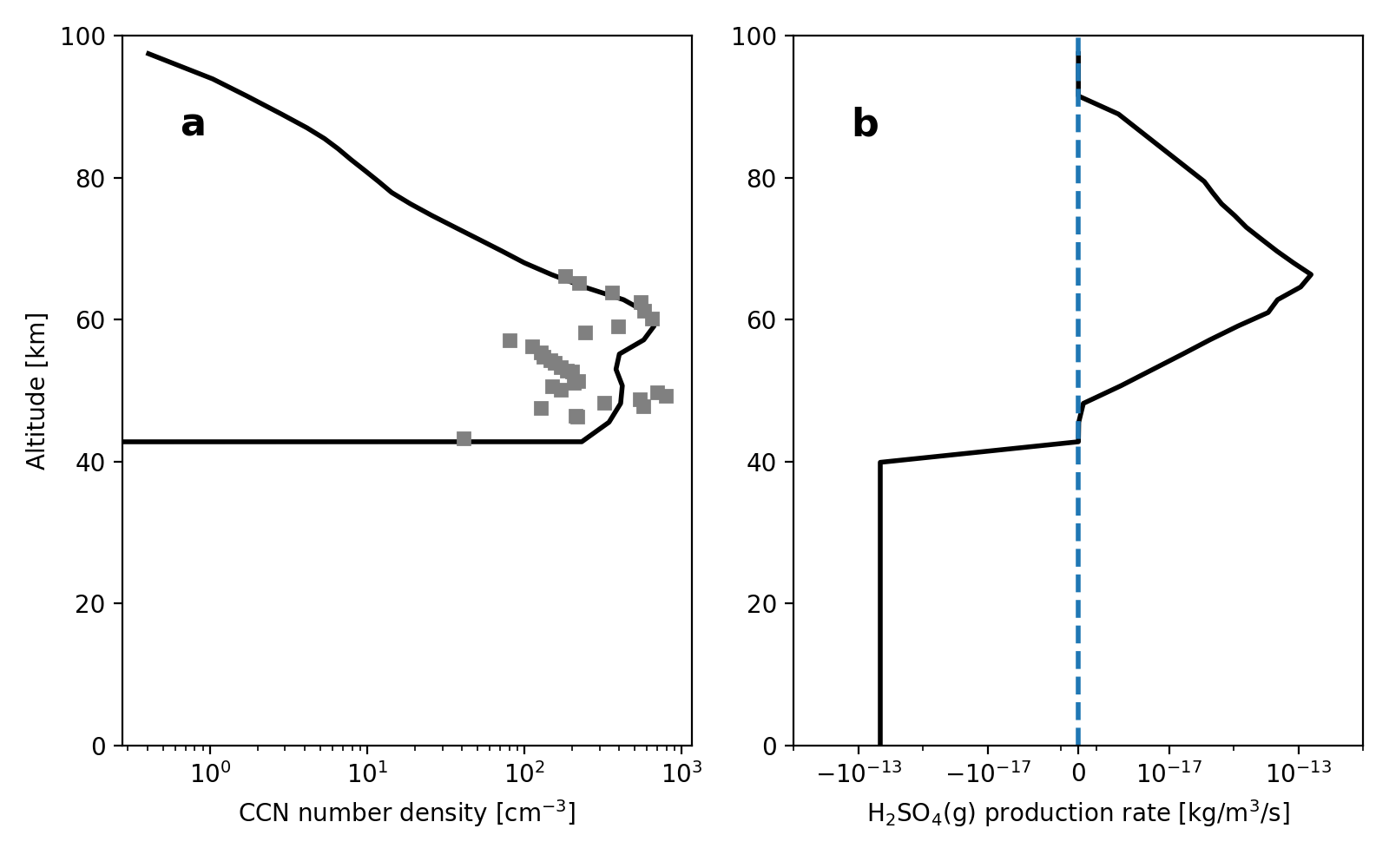}
\caption{Vertical profiles of (a) cloud condensation nucleus (CCN) number density and (b) average sulfuric acid vapor production rate used in our model. The CCN number density profile is from \mbox{\citeA{gao2014bimodal}} and is comparable to Pioneer Venus observations \cite{knollenberg1980microphysics}. The average production rate above $40\,$km is from \citeA{krasnopolsky2012photochemical}. The decomposition rate below $40\,$km is constant and chosen to balance the production above $40\,$km.}
\label{fig1}
\end{figure}

The eddy diffusion profile in \citeA{dai_simple_2022} is not used here since our model has both horizontal and vertical transport for cloud tracers driven by advection \cite{mendoncca2022mass}. The contribution from sub-grid (small-scale) processes like wave breaking and turbulence is not considered here. Following \citeA{dai_simple_2022}, we fix the cloud particle number density profile as that in \citeA{gao2014bimodal} (Fig.~\ref{fig1}), which roughly agrees with the observations from Pioneer Venus by \citeA{knollenberg1980microphysics}. We apply this profile everywhere in our 3D model. Then the average radius of the cloud particles can be calculated at each grid from the cloud mass density our model calculated and the fixed cloud particle number density. This average radius is needed for the computation of condensation rates (formula 1-2). Note that we do not treat detailed microphysics and do not resolve the size distribution of cloud particles. Similar to the 1D work by \citeA{dai_simple_2022}, this simple particle size treatment could produce micron-size particles that generally agree with mode-2 particle size observed by Pioneer Venus \cite{knollenberg1980microphysics} in the lower cloud but are slightly smaller than the observed mode-2 particles in the upper cloud (Fig.S5).

For chemistry, we adopt a prescribed average \ce{H2SO4} chemical production rate profile for the altitudes above $40\,$km (Fig.~\ref{fig1}). The solar flux reaching the top of the atmosphere is dependent on the solar zenith angle and can affect the atomic oxygen production in the Venus mesosphere that is crucial for \ce{SO3} production and thus \ce{H2SO4} vapor chemical production \cite{shao2020revisiting}. Therefore, we scale the \ce{H2SO4} production rate profile based on the cosine of the solar zenith angle. On the nightside, there is no sunlight, and thus we assume no \ce{H2SO4} chemical production. The global average of our scaled production rate profiles equals that of \citeA{krasnopolsky2012photochemical}. Below $40\,$km, to account for a full hydrogen cycle, we set up a constant \ce{H2SO4} decomposition rate profile such that the whole column chemical conversion rate between \ce{H2SO4} and \ce{H2O} is zero. For the whole atmosphere, we assume the major sink (source) of \ce{H2O} vapor is \ce{H2SO4} production (decomposition), for simplicity and also to conserve \ce{H} atoms.  

For simplicity, we use the explicit integration method, the Euler forward method, to solve the cloud-related processes (condensation/evaporation, \ce{H2SO4} production/decomposition and sedimentation) separately from the advective transport. The cloud tracers are currently passive tracers, i.e., no radiative feedback; the cloud structure for radiative calculations is fixed as in \citeA{mendoncca2020modelling}. The time step for the dynamical transport is 15 seconds in our simulations, and this time step is sufficient to reproduce the main dynamical feature of the Venus atmosphere (including super-rotation and thermal tides). To ensure numerical accuracy, the time step for the cloud physics solver is set to 1/20th of the time step for the dynamical transport. This is because typically condensation occurs at a faster rate than transport (Fig.~\ref{figtimescale}). 

We use the same settings as \citeA{mendoncca2020modelling}. The icosahedral grid level is 5, corresponding to a horizontal resolution of about 2 degrees. Vertically, we have 49 non-evenly-spaced levels from the ground to 100 km (about 2 km vertical resolution). 
One Venus solar day is about 117 Earth days. We start our simulation (with the cloud-related processes included) from the results of \citeA{mendoncca2020modelling} that has integrated the dynamical simulations for 25 000 Earth days and has reproduced the observed temperature structure and wind features like super-rotation in the cloud region. However, we note that our dynamical output is somewhat different from the previous OASIS results by  \citeA{mendoncca2020modelling} . This is mainly presented in stronger zonal wind estimated around the cloud top around the equator (Fig.~\ref{new_fig3}a; also see Section 3.2). The reason can be that we have updated OASIS for several aspects including using hydrostatic assumption and improving radiative transfer. Despite overestimated cloud-top zonal wind, the horizontal mixing of cloud tracers at the cloud top is expected to be very efficient and not significantly different from previous studies. For initial abundances of cloud tracers (\mbox{\ce{H2SO4} vapor, \ce{H2O} vapor, \ce{H2SO4} liquid and \ce{H2O} liquid}, we adopt results from \mbox{\citeA{dai_simple_2022}} and apply them to all spatial locations.

We run the new simulation for \change{1 170}{1170} Earth days (10 Venus solar days), and the results are in quasi-steady state (the results do not change over different Venus solar days). We have selected the final solar day of Venus to analyze the findings of this work. The time interval between outputs is 15 000 seconds.

\section{Results}
\label{sec3}
\subsection{Global average of the Venus cloud profiles}
\begin{figure}
\noindent\includegraphics[width=1.0\columnwidth]{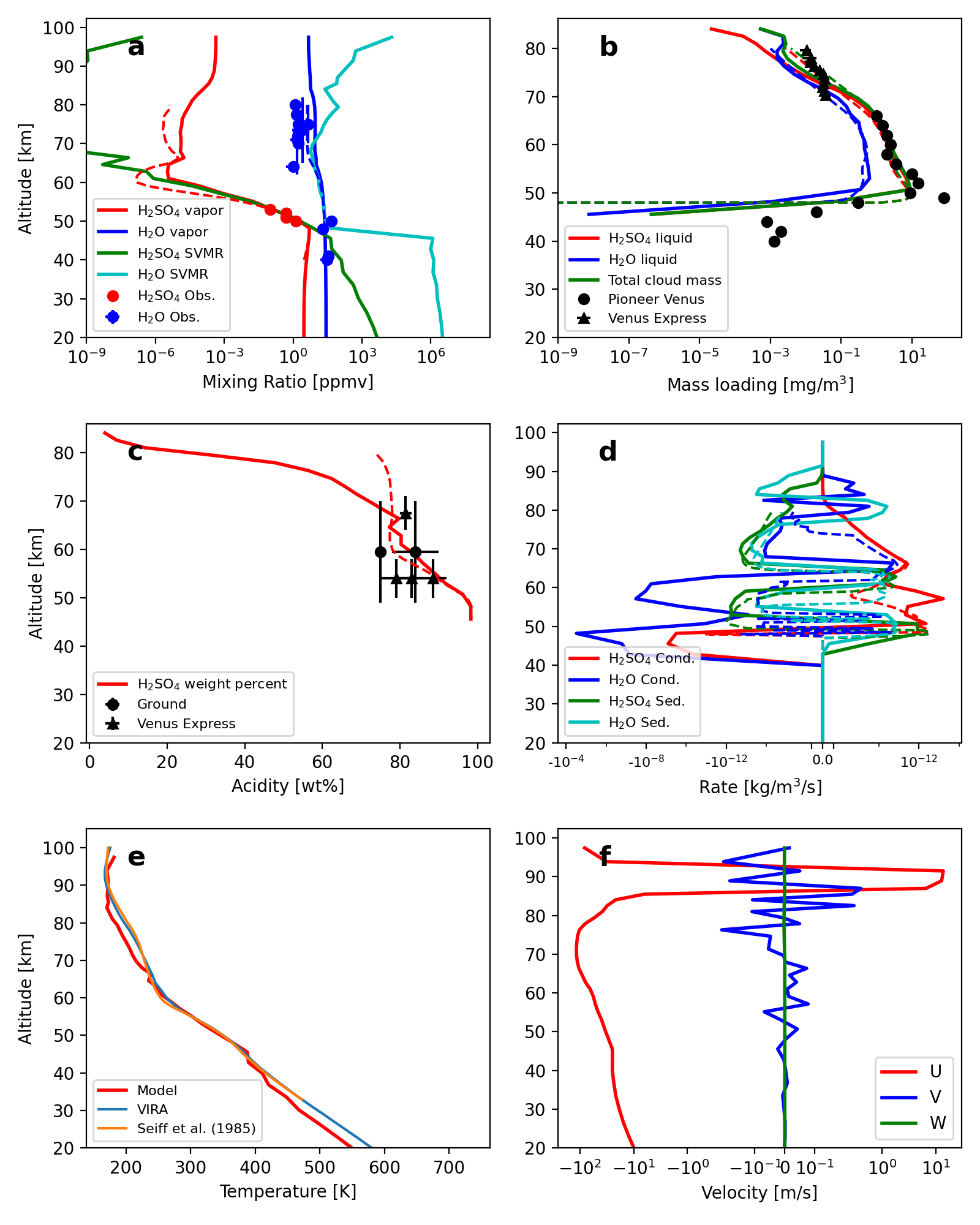}
\caption{Profiles of simulated (a) \ce{H2SO4} and \ce{H2O} vapors (units: ppmv), (b) cloud mass loading (unit: mg$\,$m$^{-3}$), (c) \ce{H2SO4} weight percentage (unit: \%), (d) condensation and sedimentation rates (units: kg$\,$m$^{-3}\,$s$^{-1}$), (e) temperature (unit: K), and (f) wind velocities (units: m$\,$s$^{-1}$) averaged globally and over the last Venus day. Dashed lines are results from \citeA{dai_simple_2022}. In panel a, the vapor observations, shown by error bars, are from \citeA{oschlisniok2012microwave}, \citeA{de1991deuterium}, \citeA{de1995water}, \citeA{meadows1996ground}, \citeA{sandor2005water}, \citeA{bertaux2007warm}, \citeA{gurwell2007swas}, \citeA{fedorova2008hdo}, \citeA{encrenaz2013hdo} and \citeA{encrenaz2020hdo}. In panel b, the cloud mass observations, shown by error bars, are from \citeA{knollenberg1980microphysics} and \citeA{wilquet2009preliminary}. Note that the mass loading from \citeA{wilquet2009preliminary} is estimated from assumed size distribution and not a direct observation. In panel c, the \ce{H2SO4} weight percent observations, shown by error bars, are from \citeA{hansen1974interpretation}, \citeA{pollack1978properties}, \citeA{barstow2012models}, \cite{cottini2012water}, \citeA{arney2014spatially} and \cite{mcgouldrick2021using}. In panel e, the light blue line is the temperature profile from VIRA (Venus International Reference Atmosphere) for $0-30^\circ$ latitude \cite{1985AdSpR...5k...1K}; the orange line is the Pioneer Venus observation at $45^\circ$ latitude \cite{seiff1985models}.}
\label{fig2}
\end{figure}

The globally mean vertical profiles of \ce{H2SO4} vapor, cloud mass loading and cloud acidity (\ce{H2SO4} weight percentage), averaged over the last Venus solar day, are in good agreement with the observations from various instruments (Fig.~\ref{fig2}a-c). Our model produces a cloud base at around 48 km, consistent with 1D models by \citeA{krasnopolsky2015vertical} and \citeA{dai_simple_2022}, but the 3D Venus PCM by \citeA{stolzenbach2023three} produces a lower cloud base (about 42 km) likely caused by cold bias in the model. Similarly, \ce{H2SO4} abruptly decreases at the cloud base in our model and in \citeA{dai_simple_2022}, but at a lower altitude in \citeA{stolzenbach2023three}. The supersaturation of \ce{H2SO4} vapor above 60 km (Fig.~\ref{fig2}a), produced in \citeA{dai_simple_2022}, is also present in our simulation. This supersaturation does not exist in the 3D model of \citeA{stolzenbach2023three} because their model uses an equilibrium cloud scheme. Cloud mass loading reaches nearly 10 mg$\,$m$^{-3}$, same as \citeA{dai_simple_2022}, and agrees with Pioneer Venus observations \cite{knollenberg1980microphysics} (Fig.~\ref{fig2}b). Acidity in the main cloud deck (47-70 km) varies between 73-98 \%, consistent with observations (Fig.~\ref{fig2}c) and 1D models \cite{krasnopolsky2015vertical, dai_simple_2022} and the Venus PCM \cite{stolzenbach2023three}. These agreements are inherited from the cloud scheme of \citeA{dai_simple_2022}, and the underlying physics for these profiles is described as the combination of condensation-evaporation, eddy diffusion and \ce{H2SO4} chemical production processes. In Fig.~\ref{fig2}d, \ce{H2SO4} condensation happens in almost the whole cloud deck region ($47-70\,$km), and evaporation mainly happens around the cloud base (Fig.~\ref{fig2}d). This rate profile is similar to that at \citeA{dai_simple_2022}. The simulated temperature \change{and winds are}{is} the same as \citeA{mendoncca2020modelling} and agree with observations \cite{seiff1985models, 1985AdSpR...5k...1K} (Fig.~\ref{fig2}e).

\begin{figure}
\noindent\includegraphics[width=1.0\columnwidth]{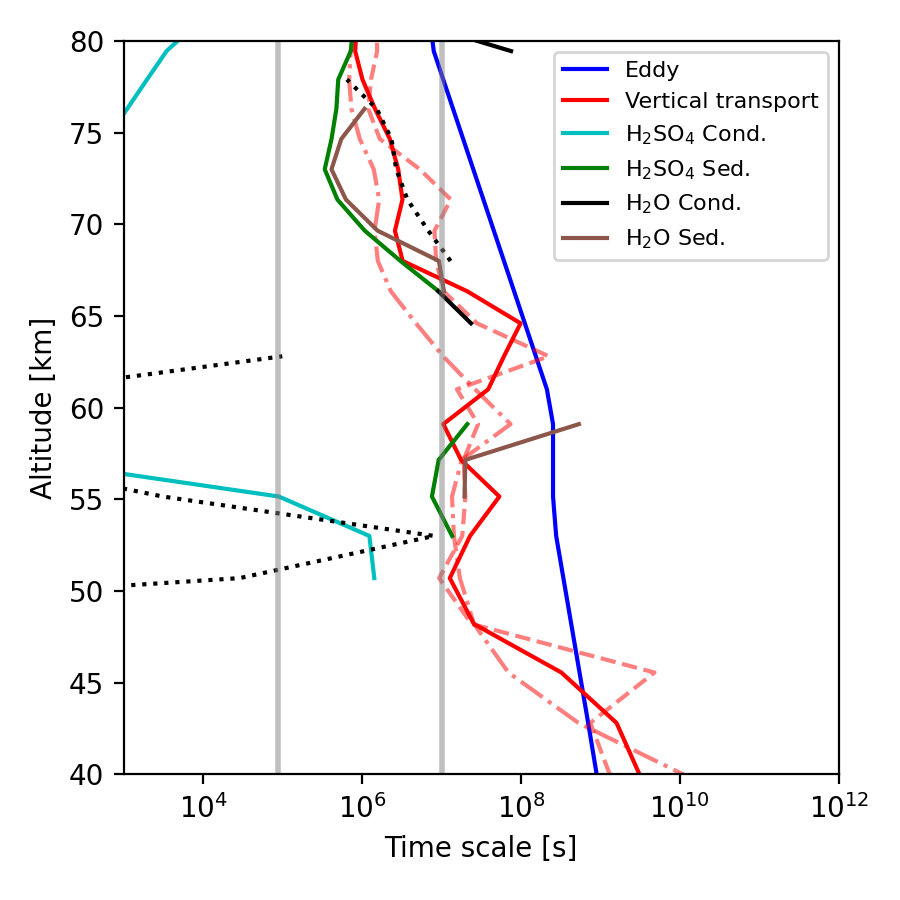}
\caption{Estimated timescales of vertical transport (red) around the equator, \ce{H2SO4} condensation (cyan), \ce{H2SO4} sedimentation (green), \ce{H2O} condensation (black), \ce{H2O} sedimentation (brown) from our model. Blue line is the time scale for eddy diffusion used in \citeA{dai_simple_2022}, estimated as $H^2/K_{zz}$. H is scale height, and $K_{zz}$ is eddy diffusion.  For vertical transport, timescale is estimated as $H/w$, where w is vertical velocity. Transparent red lines are vertical transport timescales estimated at $45^\circ$ (dashed) and $75^\circ$ (dashdot) latitudes.  For condensation and semdimentation, time scales are estimated as $\rho_i/S$, where $\rho_i$ is the mass density of species i, and S is condensation or sedimentation. Time scales for negative condensation (i.e., evaporation) are shown by dotted lines with the same color as that for positive condensation. One Earth solar day and one Venus solar day are shown by grey vertical lines.}
\label{figtimescale}
\end{figure}

However, our simulated \ce{H2O} vapor above the clouds is larger than the observations, \citeA{dai_simple_2022} and \citeA{stolzenbach2023three}, reaching $4-10\,$ppm in our model (Fig.\ref{fig2}a). This is related to the stronger vertical transport in the cloud region in our simulations. Fig.~\ref{figtimescale} shows the timescale estimated for various processes in our model. The timescale for vertical transport in the cloud region is mostly order-of-magnitude shorter than that of the eddy diffusion used in \citeA{dai_simple_2022}. Due to this strong transport, more \ce{H2O} is transported from the cloud region to above it. This effect from cloud transport effeciency is studied in \mbox{\citeA{karyu2024one}}, where double eddy diffusion in the clouds in a microphysical model can produce overestimation of \mbox{\ce{H2O}} vapor above the clouds.  The strong vertical transport could come from strong atmospheric thermal tides in our model, \add{which will be shown in Section 3.3}. 

\subsection{Zonal-mean meridional distribution of the Venus clouds}

\begin{figure}
\noindent\includegraphics[width=1.0\columnwidth]{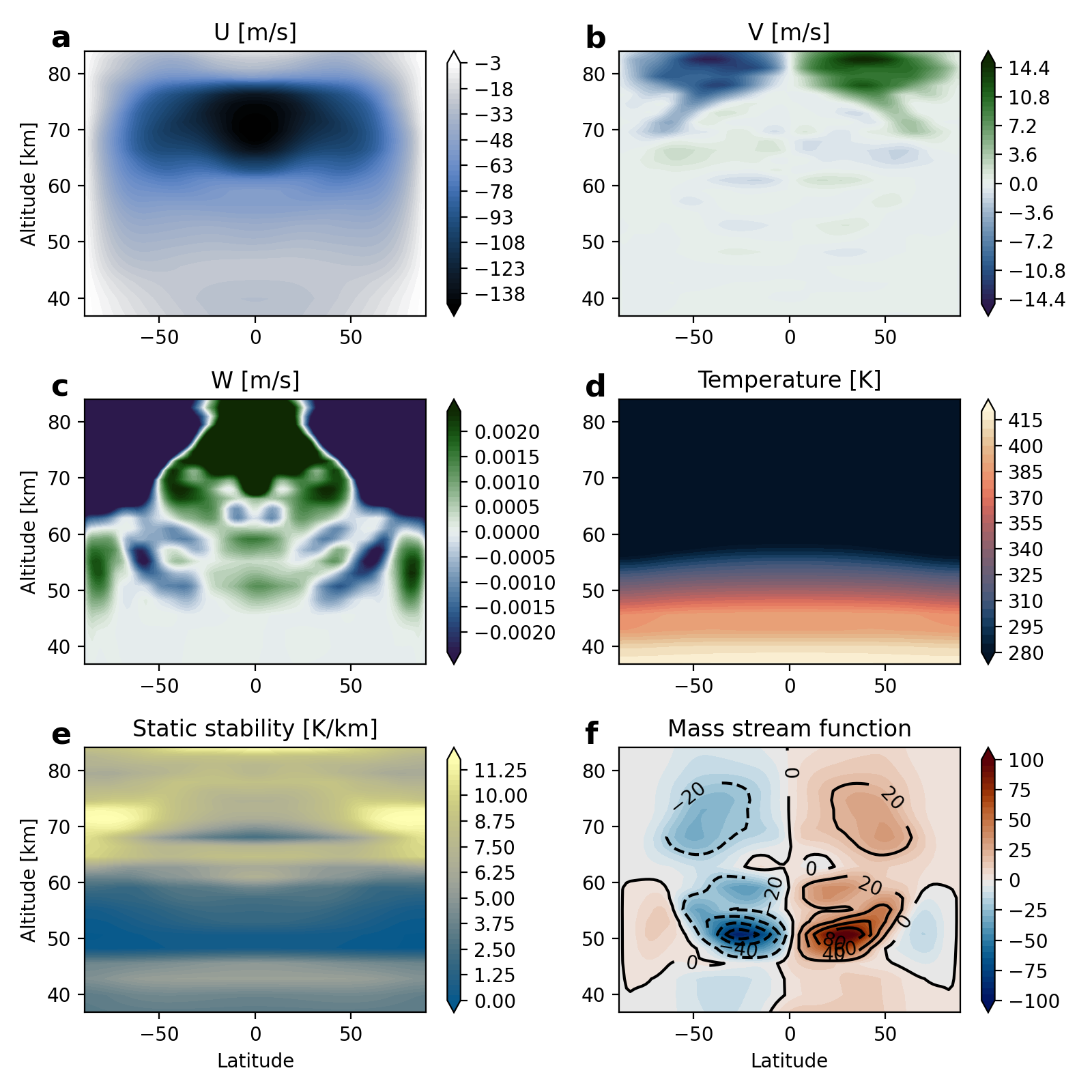}
\caption{Simulated (a) zonal wind (unit: m/s), (b) meridional wind (units: m/s), (c) vertical wind (unit: m/s), (d)  temperature (unit: K), (e) static stability (unit: K/m), (f) mass stream function (unit: $10^9$ kg/s) averaged zonally and over the last Venus day. \add{In panel f, both contour line and color represent mass stream function.} }
\label{new_fig3}
\end{figure}

In the zonally averaged plane, the Hadley-type circulation is the main feature of the region at 60-80 km (Fig.~\ref{new_fig3}f). The zonal-mean zonal wind at the equator  at 65-75 km ranges at $100-140$ m/s. This is comparable but somewhat higher than previous models \cite{lebonnois2010superrotation, mendoncca2020modelling, yamamoto2019solar}  and observations from Akatsuki and other space missions \cite<e.g.>{sanchez2017atmospheric, yamamoto2019solar}. The poleward meridional winds locate around 80 km are about 10 m/s, which are stronger than the equatorward winds in the cloud region (around 65 km); \add{above 65 km} at low latitudes, there is an ascending branch, while at high latitudes the air descends (Fig.~\ref{new_fig3}a-c). These wind fields \add{overall} agree well with other Venus GCM simulations \cite{lebonnois2010superrotation, 2016Icar..278...38L, ando2020venusian}.  Static stability is positive above 60 km (Fig.~\ref{new_fig3}e). In the low and middle clouds, the static stability approaches zero, suggesting convection is likely occurring in this region. In the upper cloud, the static stability is lower at low latitudes due to the effect of solar heating. This pattern is comparable to Venus Express and Akatsuki observations by  \citeA{ando2020thermal}. The  meridional circulation (RMMC) has two components in the middle and low clouds: a stronger one at low-middle latitudes and a weaker one at high-latitudes (Fig.~\ref{new_fig3}f). Around the cloud top (70 km), tracers will be transported upward first and then to high latitudes.

\begin{figure}
\noindent\includegraphics[width=1.0\columnwidth]{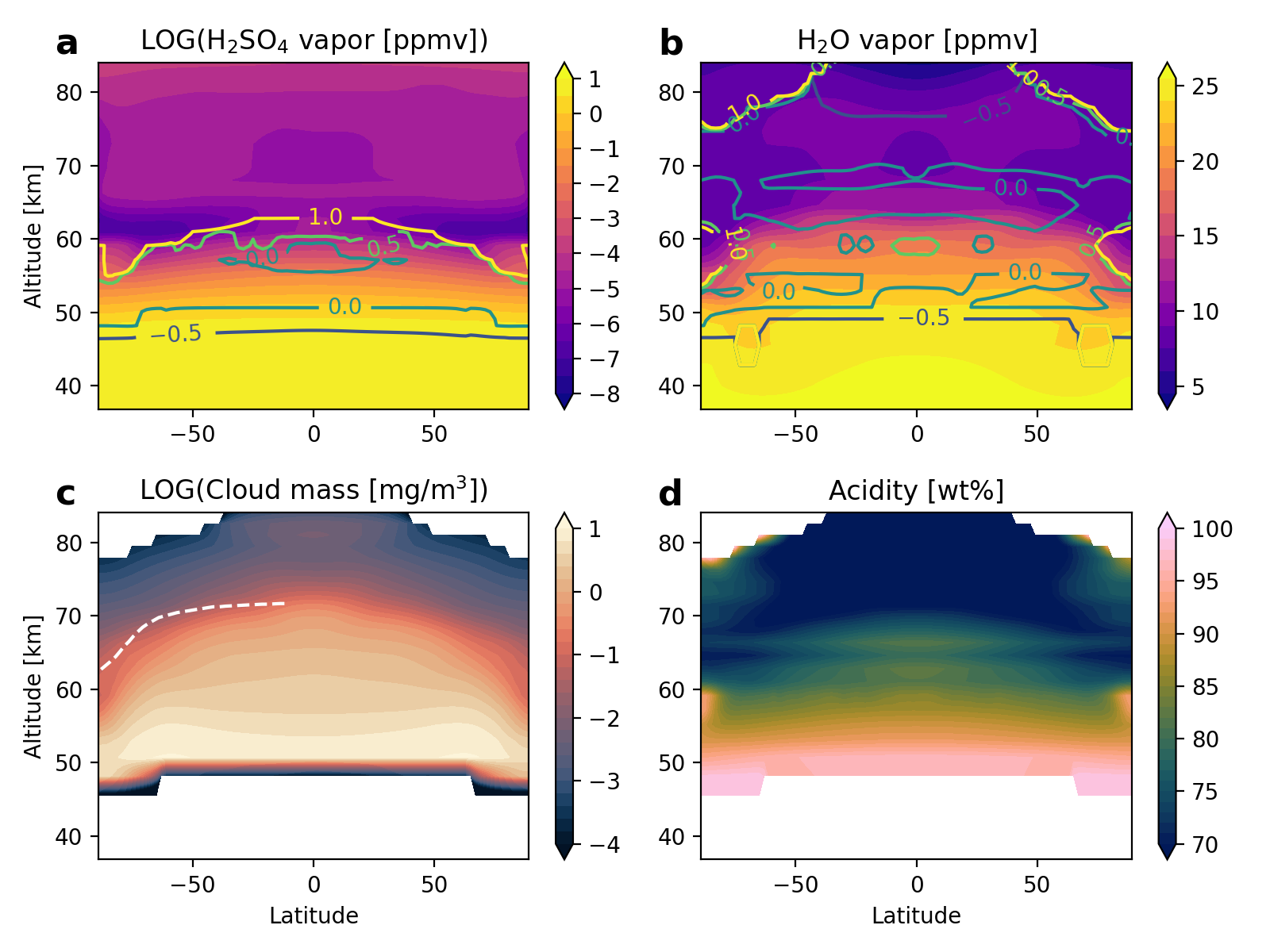}
\caption{Simulated (a) \ce{H2SO4} vapor (unit: ppmv), (b) \ce{H2O} vapor (units: ppmv), (c) cloud mass loading (unit: mg$\,$m$^{-3}$), (d) cloud acidity (unit: \%) averaged zonally and over the last Venus day. In panel a and b, saturation status is calculated as the relative difference between vapor mixing ration and saturation vapor mixing ratio (SVMR) and shown by solid lines. In panel c, the cloud top determined by the $1.5\,$um CO2 band from Venus Express \cite{ignatiev2009altimetry, cottini2012water} is shown by the white dashed line.}
\label{fig3}
\end{figure}

In our simulations, the volume mixing ratio (vmr) of \ce{H2SO4} vapor is a few ppm at 40-55 km (Fig.~\ref{fig3}a and S7a). The simulated zonally averaged meridional distribution qualitatively agrees with the \ce{H2SO4} vapor observations by \cite{oschlisniok2021sulfuric}. In the observations, there are two local enhancement of \mbox{H2SO4} vapor: one  at the equatorial region, and one at the polar regions . The enhanced region around the equator is located at around 47 km, which is also reproduced by our model. However, in the polar region, the region is at around 43 km, while our model simulates one at 45-46 km. Quantitatively, the mixing ratio values in both the enhanced regions simulated by our model (about 5 ppm) are smaller than those observed (at an order of 10 ppm). This is possibly related to cloud mass distribution near cloud base (Fig.~\ref{fig3}c; see also Fig.S7c) since clouds near the base can precipitate and induce a local enhancement of vapors, as suggested by \cite{ando2020venusian}. The meridional circulation (Fig.~\ref{new_fig3}f; see also Fig.S6) and sedimentation (Fig.S1) in our model accumulate the clouds at mid-high latitude, while in \citeA{ando2020venusian} the clouds are accumulated near the poles.  

In the main cloud deck, both volume mixing ratios of \ce{H2SO4} and \ce{H2O} vapors decreases as latitude increases (Fig.~\ref{fig3}), along with an overall temperature decrease over latitudinal dimension (Fig.~\ref{new_fig3}d). This suggests the temporature's control on both vapors' latitudinal distributions: \ce{H2SO4} at 50-60 km and \ce{H2O} in the main cloud deck basically follow their SVMRs (Fig.~\ref{fig3}a-b contour lines) that are a function of temperature. However, physics for the upper-cloud \mbox{\ce{H2SO4}} vapor are different. \ce{H2SO4} vapor above 60 km is supersaturated. This is because the supersaturation is required to produce a sufficiently high condensation rate to balance with \ce{H2SO4} chemical production rate \cite{dai_simple_2022}. Thus, there is a transition between near-saturation to supersaturation. This transition causes a minimum of \ce{H2SO4} vapor in the vertical direction (also see \citeA{dai_simple_2022}). Due to the variation of temperature vertical profiles with latitude, the vertical minimum location of \ce{H2SO4} vapor is higher at low latitudes.  Overall, both vapors' meridional distributions are significantly affected by temperature. 

Even though \ce{H2O} vapor does have some meridional variations, these variations do not exceed a factor of 2. In other words, \ce{H2O} vapor is relatively uniformly distributed over the meridional direction. This uniformity in the upper cloud is consistent with the ground-based observations by TEXES \cite{encrenaz2023hdo, encrenaz2013hdo, encrenaz2012hdo, encrenaz2016hdo, encrenaz2019hdo, encrenaz2020hdo}. The uniform distribution above the clouds is consistent with the long chemical lifetime of \ce{H2O} vapor suggested by previous models \cite{zhang2012sulfur, bierson2020chemical, shao2022local, stolzenbach2023three}. Inside the cloud region, condensation/evaporation  dominates \ce{H2O} vapor and drives it to follow its SVMR (Fig.~\ref{fig3}b). SVMR, as a function of temperature and acidity, does not vary by orders of magnitude over latitudes. Consequently, small meridional variations of \mbox{\ce{H2O}} vapor are also presented in the cloud region.

The cloud top extends higher in the lower latitudes, and this feature is qualitatively consistent with Venus Express observations \cite<e.g.,>{ignatiev2009altimetry, cottini2012water}. The Hadley-type circulation and the \ce{H2SO4} chemical production that is dependent on solar zenith angle drive this pattern. The ascending air of meridional circulations at the equatorial region extends clouds to higher altitudes. In the upper cloud, more production of \ce{H2SO4} vapor at low latitudes also causes more condensation. Similar mechanisms are also discussed by \mbox{\citeA{imamura1998venus}} and \mbox{\citeA{karyu2023vertical}}.Consequently, a meridionally-decreasing cloud top structure is produced. However, for the cloud base structure, temperature plays an important role. In the polar region, the cloud base is located at a lower altitude. This is due to the fact that the polar region has lower temperatures, resulting in a lower SVMR for \ce{H2SO4}. Thus, the mixing ratio of \ce{H2SO4} vapor can reach the saturation mixing ratio at a lower altitude. The cloud base altitude does not vary smoothly with latitude, which is possibly due to the low vertical resolution (about 2 km in our simulations). 

In the upper cloud, the cloud mass peaks at the equator in the meridional direction, while it peaks at mid-high latitudes meridionally in the low and middle clouds (Fig.~\ref{fig3}c). In the upper cloud, a weak meridional circulation tends to carry the clouds to high latitudes (Fig.~\ref{new_fig3}f). However, this does not overcome the pattern by more cloud condensation at low latitudes due to more chemical production of \mbox{\ce{H2SO4}} vapor. In the low and middle clouds, mass transport of meridional circulation is stronger, and little \mbox{\ce{H2SO4}} vapor is chemically produced here. Consequently, clouds form at the ascending motion of meridional circulation at low latitudes and are transported to mid-high latitudes by the circulation. Different from previous models \cite{imamura1998venus, ando2020venusian, stolzenbach2023three, karyu2023vertical} , our simulation also produces a weak Ferrel-like circulation in this region (Fig.~\ref{new_fig3}f). This circulation and Hadley-like circulation contributes to the transport of cloud mass to mid-high latitudes.  As a result,  the column cloud mass slightly increases from low to mid-high latitudes and decreases by about a factor of 2 at the poles (not shown here). This pattern is inconsistent with observations by Venus Express \mbox{\cite{haus2014atmospheric}}, where column cloud mass peaks at low and high latitudes. Some previous models \cite{imamura1998venus, ando2020venusian, karyu2023vertical}  have reproduced this feature but with small differences. The discrepancy of cloumn cloud mass latitudinal variation between our model and the previous models should be related to different meridional circulation patterns simulated in the cloud deck.

In the cloud deck, cloud acidity does not vary significantly in the meridional direction. The equatorial clouds are overall slightly more acidic (larger \ce{H2SO4} weight percent) than that at middle latitudes. As approaching the poles, acidity becomes slightly higher again. Around $70\,$km, cloud acidity exhibits a pattern similar to the "cold-collar" pattern \cite{taylor1980structure} in the temperature field (Fig.~\ref{fig3}). In our simulations, there is a strong positive correlation between cloud acidity and temperature (see Fig.~\ref{fig3}, ~\ref{fig4}, and ~\ref{fig6}). This may be due to the different dependences of \ce{H2SO4} and \ce{H2O} SVMRs on temperature. This correlation is also found in \mbox{\citeA{steele1981effects}}, but there \mbox{\ce{H2SO4}} liquid mass is fixed, different from our simulations where it is variable. However, this may suggest that the \mbox{\ce{H2O}} SVMR is more sensitive to temperature than the \mbox{\ce{H2SO4}} SVMR.   Futhermore, our model has no cloud feedback on radiative transfer and dynamics. For instance, cloud acidity could change particle radiative properties and affect atmospheric temperature. Once the cloud feedback is resolved, the correlation between temperature and acidity may be affected. 

Above 70 km, cloud acidity decreases rapidly with altitude. This is not consistent with previous models \mbox{\cite<e.g.>{dai_simple_2022, stolzenbach2023three}}. This may be caused by \mbox{\ce{H2O}} vapor delivered by convection from the cloud deck. It condenses and dilutes liquid particles above the clouds, leading to a low acidity \mbox{(Fig.~\ref{fig2}c and Fig.~\ref{fig3}d)}. However, this needs to be allocated little attention on since our model simulates cloud mass less than $10^{-2}$ mg m$^{-3}$ \mbox{(Fig.~\ref{fig3}c)}. This means that a small variation of either \mbox{\ce{H2O}} liquid or \mbox{\ce{H2SO4}} liquid could greatly change acidity. Our model does not have detailed microphysics, and this may cause it unable to reproduce hazes above the clouds.

The zonal distributions of \ce{H2SO4} vapor, \ce{H2O} vapor, cloud mass, and cloud acidity averaged over one Venus solar day exhibit uniform features with slight variations (Fig.S2-3). This is because our simulations do not account for the longitudinal difference of the Venus surface properties (topography, emissivity, etc.). For instance, stationary orographic waves may affect advection and thus tracer transport in the clouds \cite{bertaux2016influence}. This effect may produce longitudinal variations of cloud properties. 

\subsection{Local-Time structure of the Venus clouds}
\begin{figure}
\noindent\includegraphics[width=1.0\columnwidth]{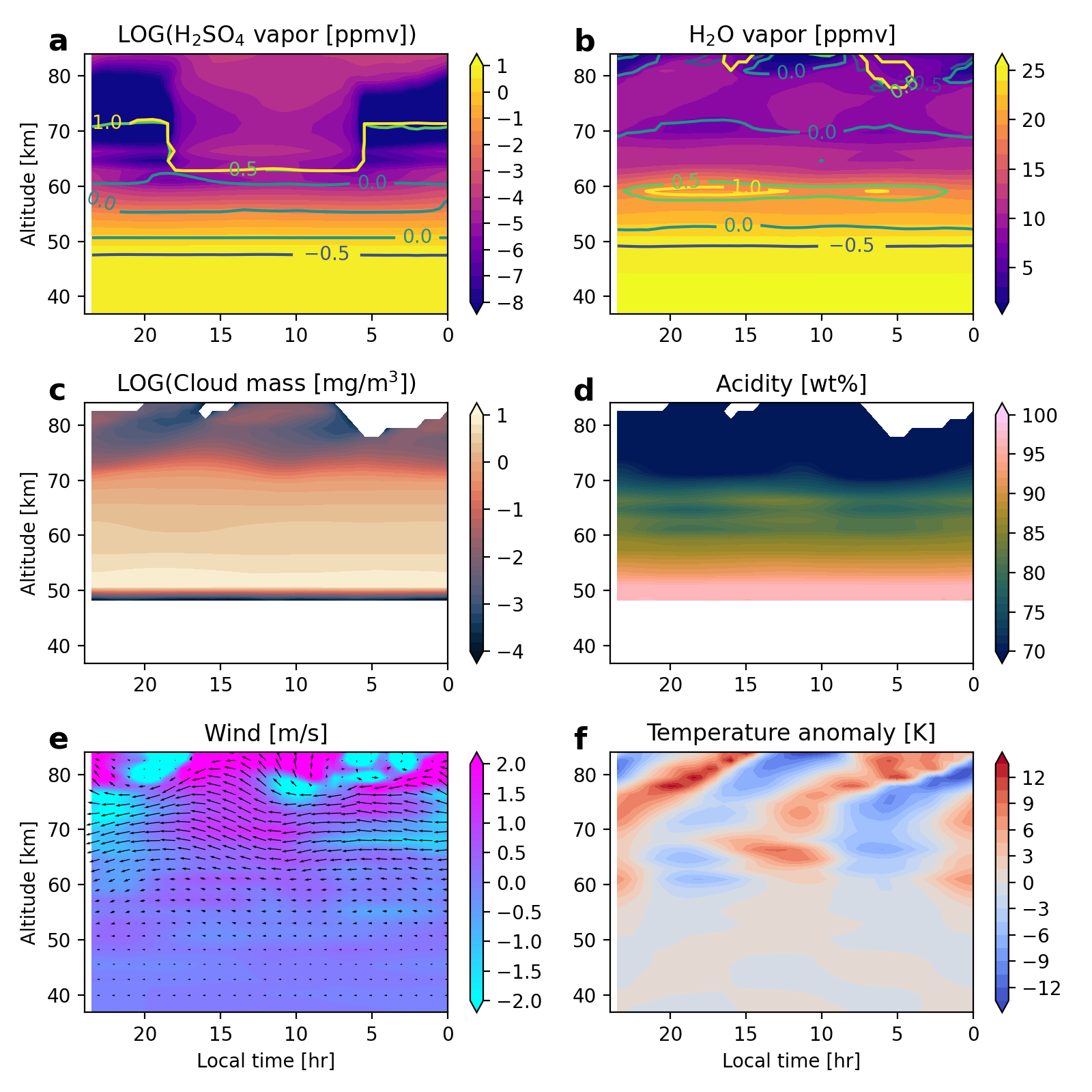}
\caption{Local-time distributions of simulated (a) \ce{H2SO4} vapor (unit: ppmv), (b) \ce{H2O} vapor (units: ppmv), (c) cloud mass loading (unit: mg$\,$m$^{-3}$), (d) cloud acidity (unit: \%), (e) wind (unit m/s) and (f) temperature anomaly (unit: K) at the equator averaged over the last Venus solar day. Morning terminator is at 6, evening terminator at 18, sub-solar point at 12, and anti-solar point at 0 or 24. In panel a and b, saturation status is calculated as the relative difference between vapor mixing ratio and saturation vapor mixing ratio (SVMR) and shown by solid lines. In panel e, color represents meridional wind; vector indicates zonal and vertical wind, and vertical wind is multiplied by 1000 for visibility.}
\label{fig4}
\end{figure}

\begin{figure}
\noindent\includegraphics[width=1.0\columnwidth]{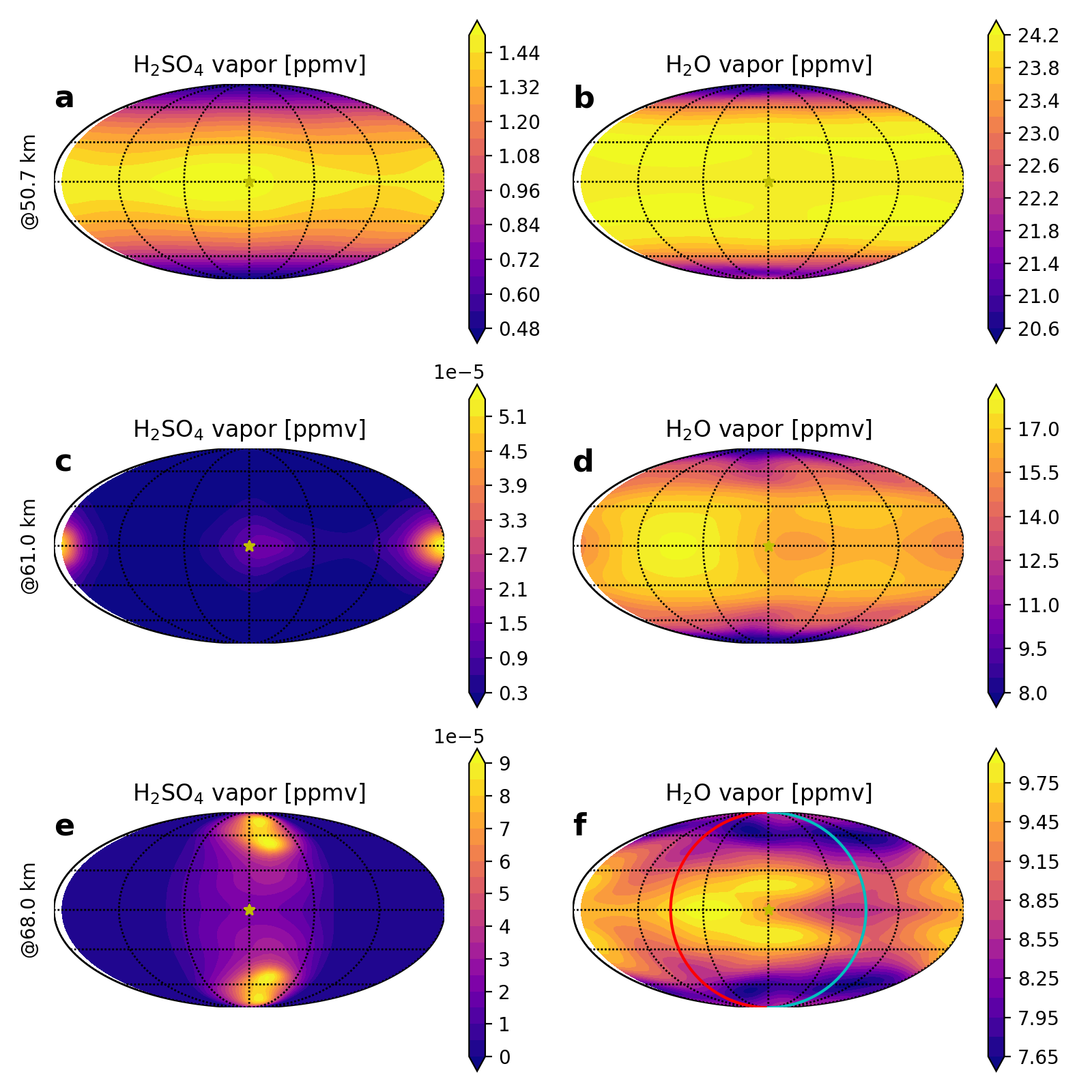}
\caption{Horizontal distributions of simulated \ce{H2SO4} vapor (unit: ppmv) (left column) and \ce{H2O} vapor (units: ppmv) (right column) at $51$ km (upper), $61$ km (middle)and $68$ km (lower) averaged over the last Venus solar day. Note that the horizontal axis is local time. The yellow star is the sub-solar point (noon or 12:00); the red and cyan lines in the last panel show the position of evening (18:00) and morning (06:00) terminators, respectively.}
\label{fig5}
\end{figure}

\begin{figure}
\noindent\includegraphics[width=1.0\columnwidth]{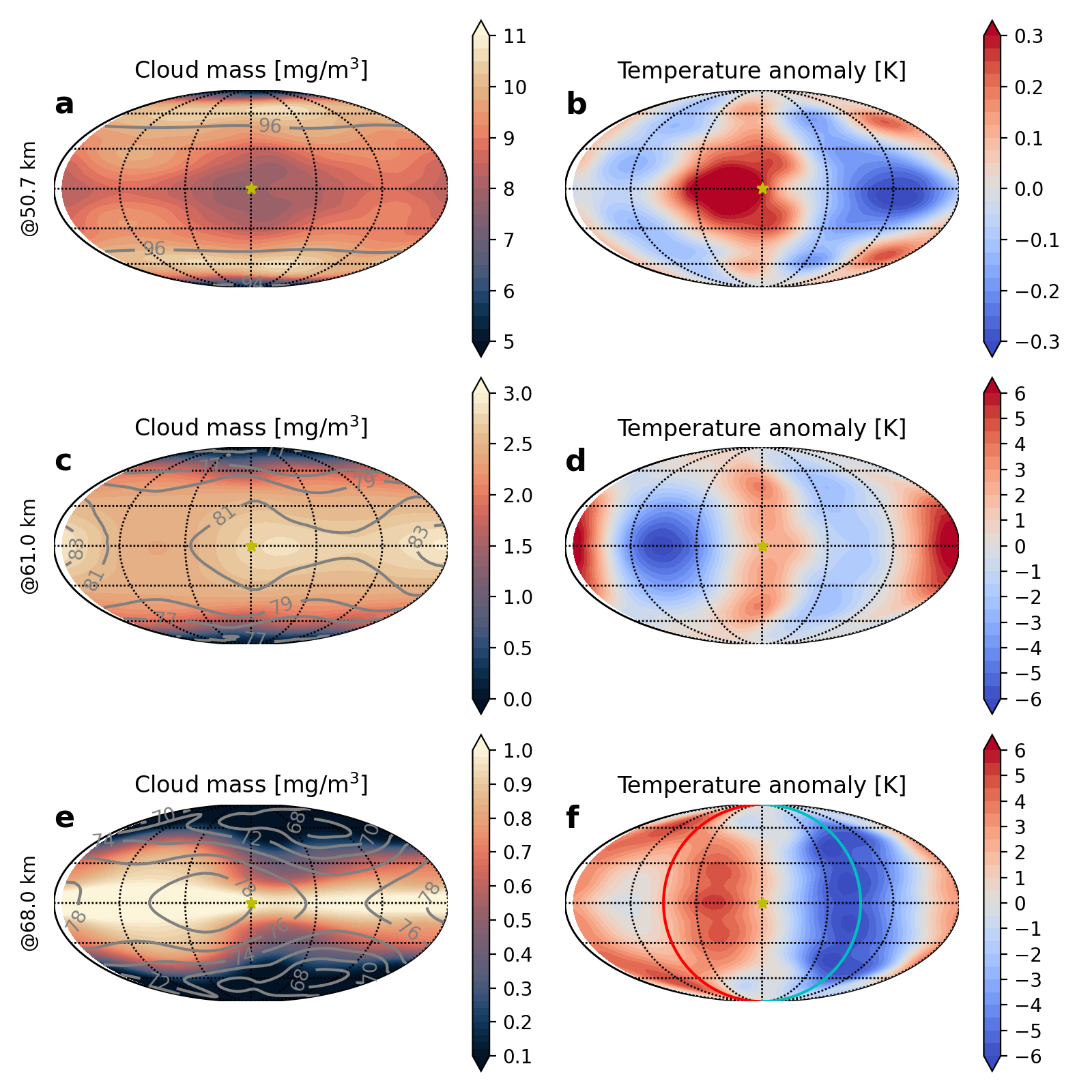}
\caption{Horizontal distributions of simulated cloud mass loading (unit: mg$\,$m$^{-3}$) (left column) and temperature anomaly from zonally averaged profiles (unit: K) (right column) at $51$ km (upper), $61$ km (middle)and $68$ km (lower) averaged over the last Venus solar day. Note that the horizontal axis is local time. In the left column, the \ce{H2SO4} weight percent contours (unit: $\%$) are shown by black lines. The yellow star is the sub-solar point (noon or 12:00); the red and cyan lines in the last panel show the position of evening (18:00) and morning (06:00) terminators, respectively.}
\label{fig6}
\end{figure}

We select local time as a coordinate and average our simulations over the last Venus solar day to obtain the local-time structure of the Venus clouds (Fig.~\ref{fig4} - ~\ref{fig6}). Overall, at $50-80$ km, \ce{H2SO4} vapor, \ce{H2O} vapor and cloud mass exhibit semidiurnal features at different altitudes at low latitudes (Fig.~\ref{fig4}a-c), which features are caused by the semidiurnal thermal tides \cite{burt1984thermal, lebonnois2010superrotation, takagi2018three, mendoncca2020modelling, 2016Icar..278...38L, mendoncca2016exploring, kouyama2019global, fukuya2021nightside}. For \ce{H2SO4} vapor, the semidiurnal feature appears significantly in the middle cloud. In this altitude range, \ce{H2SO4} vapor is controlled by condensation and follows its SVMR that is affected by temperature. Thus, the semidiurnal tide (with an amplitude of a few Kelvin, Fig.~\ref{fig6}d and f) can produce a significant semidiurnal feature of \ce{H2SO4} vapor. Below this altitude range in the low and middle clouds, \ce{H2SO4} vapor follows its SVMR, but the semidiurnal tide amplitude is small (less than 0.5 K, Fig.~\ref{fig6}a) and cannot cause large variations of \ce{H2SO4} vapor over local time. Above that altitude range, \ce{H2SO4} vapor exhibits a large day-night difference (Fig.~\ref{fig4}a and ~\ref{fig5}e). This is because \ce{H2SO4} vapor becomes supersaturated above the upper cloud and is balanced mainly by condensation and chemical production. Where the chemical production is higher, the supersaturation is higher, leading to a higher-concentration \ce{H2SO4} vapor (Fig.~\ref{fig5}e).

The semidiurnal feature of \ce{H2O} vapor is not significant in the low and middle clouds (Fig.~\ref{fig4}b and ~\ref{fig5}b). However, in the upper cloud and at 70-80 km, the semidiurnal feature becomes clear (Fig.~\ref{fig4}b and ~\ref{fig5}d-f). For example, at 61 and 68 km this feature has an amplitude of around 2-3 ppm and 1 ppm respectively. This is because, in the clouds, \ce{H2O} vapor follows its SVMR that is affected by the temperature field and thus by thermal tides. The dependence of the amplitude of semidiurnal tide on altitude (Fig.~\ref{fig6}) causes the altitude variation of the semidiurnal feature of \ce{H2O} vapor.

Cloud mass shows its significant semidiurnal feature both inside and above the clouds (Fig.~\ref{fig4} and ~\ref{fig6}) at low latitudes. This is because cloud tracers (\ce{H2SO4} and \ce{H2O} condensates) are basically long-term species and are affected by transport. The vertical transport caused by the semidiurnal thermal tides produces two peaks in the cloud mass local-time distribution at low latitudes. However, at some altitudes (e.g., $68$ km) there is a phase displacement between cloud mass and temperature maps (Fig.~\ref{fig6}e and f). This could be due to the zonal wind that can shift the cloud mass distribution (e.g., zonal wind can shift peaks of tracers westward in the upper cloud \cite{shao2022local}). \add{Cloud acidity also shows the semidiurnal feature in the upper cloud} (Fig.~\ref{fig4} and~\ref{fig6}) \add{. This may come from the combination of the semidiurnal tide effects on liquid} \ce{H2SO4} \add{and liquid} \ce{H2O} \add{abundances.}


At the cloud top, there are two peaks of \ce{H2SO4} vapor on the dayside, located at middle to high latitudes of the northern and southern hemispheres, respectively (Fig.~\ref{fig5}e). This is caused by meridional circulation (Fig.~\ref{new_fig3}). At middle to high latitudes, chemical production is weak due to solar zenith angle. In this situation, \ce{H2SO4} vapor is balanced mainly by advection and condensation. The descending motion of the meridional circulation at middle to high latitudes can bring \ce{H2SO4} vapor to the cloud top (Fig.~\ref{new_fig3}). This causes \ce{H2SO4} vapor to be more abundant at middle to high latitudes compared to the same altitude at low latitudes. This two-peak feature is not present in \citeA{stolzenbach2023three} at the cloud top, but a similar pattern is found at 62 km with different peaking latitudes in that work.

The semidiurnal thermal tide at low latitudes, with an amplitude of a few Kelvin, simulated by our model is stronger than that observed by LIR onboard Akatsuki ($1-2$ K) by \cite{kouyama2019global, fukuya2021nightside} and consistent with that observed by OIR onboard Pioneer Venus ($3-4$ K) \cite{taylor1980structure}. As pointed out by \citeA{kouyama2019global}, the observation difference could be due to the width of the contribution function. \add{The coincidence of strong thermal tides and strong zonal wind at low latitudes could suggest the contribution of thermal tides to the angular momentum of super-rotation, as shown by} \citeA{takagi2018three} and \citeA{yamamoto2019solar}. There is 4-6 hour phase difference between our simulation and the LIR/Akatsuki observation: while Akatsuki observes temperature maxima at around 9 and 20 hours for the low latitudes, our model simulations produce maxima at afternoon and midnight. Thus, the simulated cloud properties that are caused by the semidiurnal tide (e.g., Fig.\ref{fig5}f and Fig.\ref{fig6}e) could have a phase displacement compared to actual cloud local-time distributions.

\section{Discussion and Conclusions}
\label{sec4}
In this study, we developed a cloudy GCM for the Venus atmosphere. The cloud physics is from \citeA{dai_simple_2022} and can resolve cloud acidity self-consistently. Our model can well simulate the observed vertical cloud structure and show agreements in \ce{H2SO4} vapor and cloud top meridional patterns with Venus Express observations. Our simulations make predictions of the spatial and local-time distributions of the Venus clouds and highlight the complexity of the Venus atmosphere. 

The global-average vertical profiles of \ce{H2SO4} vapor, cloud mass loading and acidity are consistent with observations. \ce{H2O} vapor in the middle atmosphere is overestimated, compared to observations. This discrepancy in \ce{H2O} vapor is likely due to more efficient vertical transport in the cloud region in our model than the real atmosphere. Or including the hygroscopicity of \mbox{\ce{H2SO4}} would more efficiently remove \mbox{\ce{H2O}} vapor from the air.  

The general latitudinal distribution for \ce{H2SO4} and \ce{H2O} vapors in the cloud deck is that the vapor volume mixing ratio decreases as latitude increases. This is related to the dependence of their SVMRs on temperature. The cloud top is higher at the equator than at poles, which qualitatively agrees with the Venus Express observations \cite<e.g.,>{ignatiev2009altimetry, cottini2012water}. This feature can originate from the \ce{H2SO4} chemical production variations and meridional circulation. The upper cloud peaks around the equator due to \ce{H2SO4} production, while the middle cloud peaks at mid-high latitude due to meridional circulation and sedimentation. This middle cloud latitudinal distribution is inconsistent with observations and previous models. The inconsistency is probably due to the difference in simulated meridional circulations, which needs further investigation. The cloud base is also higher at the equator due to the temperature's control on the \ce{H2SO4} SVMR. Cloud acidity does not vary significantly with latitude but exhibits a similar pattern to temperature . This similarity needs more investigation since both temperature and acidity are important for the SVMRs of \ce{H2SO4} and \ce{H2O}. In the zonal direction, there are no significant differences between cloud properties at different longitudes when the results are averaged over one Venus solar day. The longitudinal dependence of surface properties (e.g., topography, emissivity, etc.), not resolved in our current model, can produce zonal variations of cloud properties.

At the equator, \ce{H2SO4} vapor, \ce{H2O} vapor, cloud mass and cloud acidity exhibit semidiurnal features at different altitudes. At $50-60\,$km, the \ce{H2SO4} vapor shows the semidiurnal pattern, related to temperature's impact on SVMR. Above this altitude range, it is controlled mainly by chemical production and condensation, and thus the day-night difference becomes significant. However, the mixing ratio of \ce{H2SO4} vapor above $60\,$km is very low and within the current remote sensing sensitivity limit. Its local-time dependence may thus pose a challenge for future Venus missions. \ce{H2O} vapor and cloud mass show the significant semidiurnal feature in the upper cloud. Future observations can verify this prediction. These results indicate the important role that the semidiurnal thermal tide plays to the local-time cloud distribution.

We have taken the initial step towards creating a Venus climate model that is entirely self-consistent and incorporates cloud physics. We aim to obtain a GCM with varying levels of complexity, including photochemistry, clouds and surface-atmosphere interactions. This is essential  for the preparation of future Venus missions like EnVision \cite<e.g.,>{2012Envision, widemann2020envision}. For missions like VERITAS \cite<e.g.,>{smrekar2022veritas}and DAVINCI+\cite<e.g.,>{garvin2022revealing}, this GCM can combine with observations like surface properties and atmospheric aerosol properties to improve our understanding of the Venus atmosphere.  To approach this goal, a crucial improvement for our model will be to remove the fixed cloud particle number density profile that is also globally uniform in our current simulations. If computational cost is reasonable, we will incorporate a more physically-based cloud physics scheme like \citeA{maattanen2023development} and \citeA{mcgouldrick2023influence}. Another improvement will be to include cloud radiative feedback and photochemistry. The current chemical scheme is rather simple and cannot cover the important parts of the sulfur cycle. Including the photochemistry will be a great benefit for studying surface-atmosphere mass exchange processes like volcanism. Furthermore, our current horizontal resolution (equivalently 2 degrees) or vertical resolution (about 2 km) is possibly not high enough to resolve small-scale turbulences, which may be important for tracer transport. Therefore, we will increase our resolution and explore the sensitivity of our results to it in the future.

\section{Open Research}
The data used to produce the figures in this work are available at \citeA{Shao2023}. The figures were made with Python version 3.10.12 \cite{van2009introduction} and Matplotlib version 3.6.1 \cite{Hunter:2007, thomas_a_caswell_2022_7162185}.

\acknowledgments
W.D.S. and J.M.M. acknowledge financial support from the PRODEX EnVision project number 4000136451. L.D. is supported by the National Natural Science Foundation of China (Grant 42305135) and Natural Science Foundation of Hunan Province (Grant 2023JJ40664). \texttt{OASIS} was run on the HPC cluster at the Technical University of Denmark \cite{DTU_DCC_resource}. We thank Dr. Russell Deitrick for providing some of our analysis tools. We also acknowledge the constructive comments from the reviewers.


\bibliography{VenusClouds}

\newpage
\centering{\includegraphics[scale=0.75,page=1]{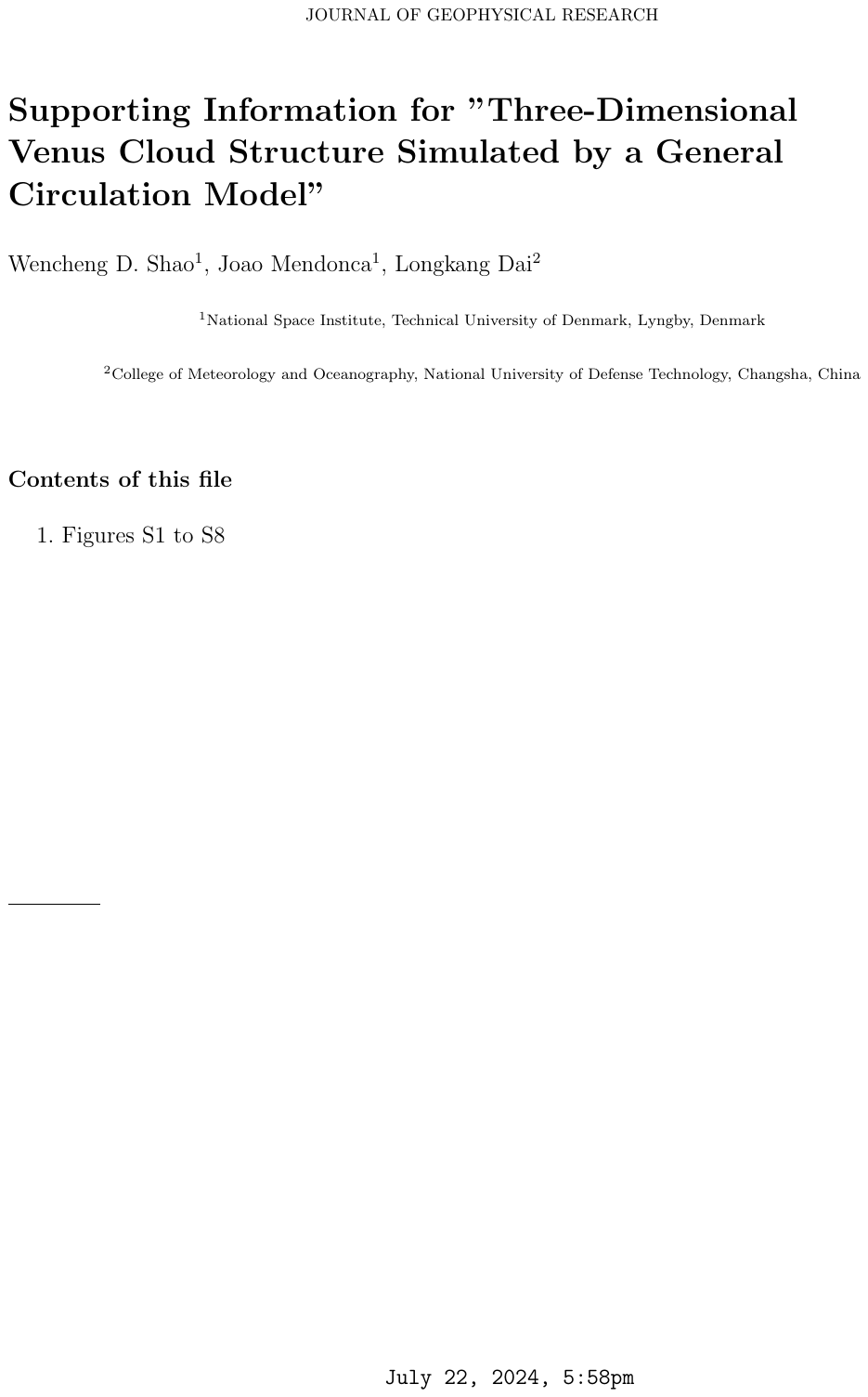}}
\newpage
\centering{\includegraphics[scale=0.75,page=2]{Shao_2023_VenusClouds_R3_SI.pdf}}
\newpage
\centering{\includegraphics[scale=0.75,page=3]{Shao_2023_VenusClouds_R3_SI.pdf}}
\newpage
\centering{\includegraphics[scale=0.75,page=4]{Shao_2023_VenusClouds_R3_SI.pdf}}
\newpage
\centering{\includegraphics[scale=0.75,page=5]{Shao_2023_VenusClouds_R3_SI.pdf}}
\newpage
\centering{\includegraphics[scale=0.75,page=6]{Shao_2023_VenusClouds_R3_SI.pdf}}
\newpage
\centering{\includegraphics[scale=0.75,page=7]{Shao_2023_VenusClouds_R3_SI.pdf}}
\newpage
\centering{\includegraphics[scale=0.75,page=8]{Shao_2023_VenusClouds_R3_SI.pdf}}
\newpage
\centering{\includegraphics[scale=0.75,page=9]{Shao_2023_VenusClouds_R3_SI.pdf}}

\end{document}